# Metal hydrogen critical temperature at the pressure 500 GPa

N. A. Kudryashov, A.A. Kutukov, E. A. Mazur [a)]

*National Research Nuclear University "MEPHI", Kashirskoe sh.31, Moscow 115409, Russia*

Éliashberg theory, being generalized to the case of the electron-phonon (EP) systems with the not constant density of electronic states, as well as to the frequency behavior of the renormalization of both the electron mass and the chemical potential, is used to study the normal properties and $T_c$ of the metal hydrogen $I41/AMD$ phase under pressure of P = 500 GPa. The phonon contribution to the anomalous electron Green's function (GF) is considered. The pairing is considered within the overall width of the electron band, and not only in a narrow layer at the Fermi surface. The frequency dependence of the electron mass renormalization $\mathrm{ReZ}(\omega)$, the density of electronic states $N(\varepsilon)$, the spectral function of the electron-phonon interaction, obtained by calculation are used to calculate the electronic abnormal GF. The Éliashberg generalized equations with the variable density of electronic states are resolved. The dependence of both the real and the imaginary parts of the order parameter on the frequency in the $I41/AMD$ phase is obtained. The $T_c = 217K$ value for the metal hydrogen $I41/AMD$ phase at the pressure P = 500 GPa has been defined.

## 1. Introduction

The metallic hydrogen is expected to have very high critical superconducting temperature $T_c$ ~ 200-400 K according to the theoretical study by Ashcroft [1]. Moreover, the calculations by Brovman and Kagan [2-5] showed that atomic hydrogen may form dynamically stable anisotropic phase of the metallic hydrogen under pressure and the metastable phase under reducing high pressure down to the atmospheric pressure. The reaching the superconducting state in the metallic hydrogen or in the hydrogen-containing compounds would be important for understanding the mechanism of high temperature superconductivity and may show a way towards the room temperature superconductivity. However, the conversion of the molecular hydrogen into a metal requires homogeneous compression under huge pressure of about 400-500 GPa which is not achievable yet. Nevertheless, the estimation of the pressure range in which the metallic phases of the hydrogen exist as well as the stability and the properties of these phases is an actual task which can be solved by ab-initio calculations. In this respect, there is a large number of studies at present [6-12]. In [13] the existence of a dynamically stable phase of the metallic hydrogen with the I41 / AMD white tin symmetry (two atoms in the primitive cell) has been theoretically predicted. In the present work we analyse I41 / AMD metal hydrogen phase at the high pressure of 500 GPa and describe the normal and the superconducting properties of the stable metallic phase with the I41 / AMD symmetry which does not contain imaginary frequencies in the phonon spectrum. This phase of the hydrogen has the properties of metallicity. The phonon spectrum of the metallic hydrogen with the I41 / AMD white tin symmetry contains no imaginary frequencies at a specified pressure [13].

**2. The theory of the electron-phonon system with the not constant density of electronic states**

The calculation of the superconducting transition temperature is usually carried out either by using extremely lax and insufficiently reasoned description of the superconductivity with a freely selectable electron density functional [14−16] without any possibility of taking into account the effect of the nonlinear nature of the Éliashberg equations either within the approximate solutions of the Éliashberg equations [17−33] not taking into account the variable nature of the electronic density of states $N(\omega)$. The solution of the Éliashberg equations on the real axis is considered to be a difficult task. Therefore the critical temperature $T_c$ within the Éliashberg formalism is usually calculated in the Green's function presentation with the use of a set of discrete Green's function values on the imaginary axis and not taking into account the variable nature of the electronic density of states. The analytic continuation of such a solution in order to determine the frequency dependence of the order parameter is extremely inaccurate. The aim of the present work is to derive the generalized Éliashberg equations for the precise $T_c$ calculation in the materials with the strong EP interaction, allowing a quantitative calculation and prediction of the normal and superconducting properties and $T_c$ in the different phases of the metal hydrogen, as well as in the high-temperature materials with the EP mechanism of superconductivity, which will be discovered in the near future. We take into account all the features of the frequency behavior of the spectral function of the EP interaction , the frequency behavior of the electron density of states $N_0(\omega)$ and the specific properties of matter in which the superconducting state is established.

For this purpose we construct in the present paper a revised version of the Migdal-Éliashberg theory for the EP system at nonzero $T \neq 0$ temperature in the Nambu representation with the account of several factors such as the variability of the electronic density of states $N_0(\varepsilon)$ within the band, the frequency and the temperature dependence of the complex mass renormalization $\operatorname{Re} Z(\omega,T)$, $\operatorname{Im} Z(\omega,T)$, the frequency and the temperature dependence of the $\operatorname{Re}\chi(\omega,T)$, $\operatorname{Im}\chi(\omega,T)$ terms usually referred to as the «complex renormalization of the chemical potential», the spectral function $\alpha^2 F$ of the electron-phonon interaction obtained by calculation, as well as the effects resulting from both the electron-hole non-equivalence and the fact of the electron band width finiteness.

Under these conditions, the derivation of the Éliashberg equations $[17-18]$ is performed anew on a more rigorous basis, leading to the new terms in the equations for the order parameter, which are not accounted for in the previous versions of the theory (see., e.g. $[17-33]$). As it has been shown in $[34]$ in the case of the strong electron-phonon interaction the reconstruction of the real part $\operatorname{Re}\Sigma$ as well as of the imaginary part $\operatorname{Im}\Sigma$ of the GF self-energy part (SP) in the materials with the variable density of electronic states is not limited to the frequency $\omega_D$ domain of the limiting phonon frequency, and extends into the much larger frequency range $\omega \gg \omega_D$ from the Fermi surface. As a result, the EP interaction modifies self-energy part of the Green's function, including its anomalous part, at a considerable distance from the Fermi surface in the units of the Debye phonon frequency $\omega_D$, and not only in the vicinity $\mu-\omega_D<\omega<\mu+\omega_D$ of the Fermi surface.

Given all that is written above, we consider the EP system with the Hamiltonian which includes the electronic component $\hat{H}_e$, the ionic component $\hat{H}_i$ and the component corresponding to the electron-ion interaction in the harmonic approximation $\hat{H}_{e-i}$, so that $\hat{H}=\hat{H}_e+\hat{H}_i+\hat{H}_{e-i}-\mu\hat{N}$. Here the following notations are introduced: $\mu$ as the chemical potential, $\hat{N}$ as the operator of electron's number in the system. The matrix electron Green's function $\hat{G}$ is defined in the representation given by Nambu as follows: $\hat{G}(x,x')=-\langle T\Psi(x)\Psi^+(x')\rangle$ , where conventional creation and annihilation operators of electrons appear as Nambu operators. Writing down the standard movement equations for the electron wave functions and averaging with the Hamiltonian $\hat{H}$ we will be able to obtain the equation for the electron GF. It is easy to show by analogy with [35] taking into account in addition to the normal and the superconducting state that the self-energy part matrix $\hat{\Sigma}(x,x')$ corresponding to the EP interaction with the vertex function $\hat{\Gamma}$ and the full account of electron-electron correlations looks like

$$\hat{\Sigma}(x,x')=-\int dx_1\int d\boldsymbol{r}''\frac{e^2}{|\boldsymbol{r}-\boldsymbol{r}''|}\varepsilon^{-1}(\boldsymbol{r}''\tau,x_1)\hat{\tau}_3\hat{G}(x,x_2)\hat{\tau}_3\hat{\Gamma}(x_2,x',x_1)+\sum_{n,\kappa;n',\kappa'}\int dx_3dx_4\times$$
$$\times\varepsilon^{-1}(x,x_3)\nabla_\alpha V_{ei\kappa}(\boldsymbol{r}_3-\boldsymbol{R}^0_{n\kappa})\varepsilon^{-1}(x_1,x_4)\nabla_\beta V_{ei\kappa}(\boldsymbol{r}_4-\boldsymbol{R}^0_{n'\kappa'})D^{\alpha\beta}_{n\kappa n'\kappa'}(\tau_3-\tau_4)\hat{\tau}_3\hat{G}(x,x_2)\hat{\tau}_3\hat{\Gamma}(x_2,x',x_1). \quad (1)$$

In (1) $x\equiv\{\boldsymbol{r},t\}$ , $\boldsymbol{R}^0_{n\kappa}=\boldsymbol{R}^0_n+\boldsymbol{\rho}_\kappa$ is the radius-vector of the equilibrium position of $\kappa$ -type ion in the crystal, $V_{ei\kappa}$ is the potential of the electron-ion interaction, with the matrix symbol $\hat{\tau}_i$ ( i=0,1,2,3) the standard Pauli matrices [27] are introduced. The phonon Green's function shall be defined as $D^{\alpha\beta}_{n\kappa,n'\kappa'}(\tau)=-\langle T_\tau(u^\alpha_{n\kappa},u^\beta_{n'\kappa'})\rangle$. Here $u^\beta_{n\kappa}$ is an $\alpha$

-projection of a deviation of $\kappa, n$ ion from the balance position $\mathbf{u}_{n\kappa} = \mathbf{R}_{n\kappa} - \mathbf{R}^0_{n\kappa}$, $\hat{\Gamma}$ is the vertex function being a matrix in $\hat{\tau}_i$ space. The vertex $\hat{\Gamma}$ behavior is supposed to be formed under the influence of the first term in (1) which incorporates all the effects of the electron-electron correlations. Hereinafter we shall not draw the first contribution $\hat{\Sigma}_{el-el}(x,x')$ from (1) having it in mind and considering via behavior of both $\hat{\Gamma}$ and $\hat{\Sigma}_{el-el}(x,x')$ all revealed earlier effects of the electron-electron correlations and electron-magnon interaction in the crystal. In (1) in accordance with the Migdal theorem[36] the vertex corrections to the second electron-phonon term should be neglected. For the self-energy part $\hat{\Sigma}(\vec{p}, i\omega_m)$ of the electron GF $\hat{g}(\vec{p}, i\omega_m)$ at the frequency points $\omega_m = (2m+1)\pi T, m = 0, \pm 1, \pm 2..$ on the imaginary axis the following decomposition on the Pauli matrixes can be written down [27]

$\hat{\Sigma}(\vec{p}, i\omega_m) = i\omega_m\left[1 - Z(\vec{p}, \omega_m)\right]\hat{\tau}_0 + \chi(\vec{p}, \omega_m)\hat{\tau}_3 + \varphi(\vec{p}, \omega)\hat{\tau}_1$, so that

$\hat{g}^{-1}(\vec{p}, i\omega_m) = i\omega_m\hat{\tau}_0 - \xi_{\vec{p}}\hat{\tau}_3 - \hat{\Sigma}(\vec{p}, i\omega_m)$. Here $\xi_{\vec{p}}$ is the electron energy not renormalized by the electron-phonon interaction. The impulses of electrons $\vec{p}$ are not supposed to lie on the Fermi surface. The integration is fulfilled in the entire Fermi volume and not just on the Fermi surface. Here $\chi(\vec{p}, \omega_m)$ is a chemical potential renormalization due to the EP interaction. For the normal state of the matter the $\chi(\vec{p}, \omega_m)$ value, which, after analytic continuation determines the frequency dependent chemical potential shift is defined by the following formula $\chi(\vec{p}, \omega_m) = \frac{1}{2}\left[\Sigma(\vec{p}, \omega_m) + \Sigma(\vec{p}, -\omega_m)\right]$. For the normal state of the matter the $Z(\vec{p}, \omega_m)$ value, which, after analytic continuation determines the mass renormalization is given by the following expression

$i\omega_m\left[1 - Z(\vec{p}, \omega_m)\right] = \frac{1}{2}\left[\Sigma(\vec{p}, \omega_m) - \Sigma(\vec{p}, -\omega_m)\right]$. Replace $Z(\vec{p}, \omega)$ by the quantity $Z(\xi, \omega)$

corresponding to the constant energy $\xi$. After the analytic continuation of $Z(\vec{p},i\omega_m)$ and $\chi(\vec{p},\omega_m)$ to the domain of the complex $\omega$ variable the functions $Z(\vec{p},\omega)$ and $\chi(\vec{p},\omega)$ become even and complex for all frequency $\omega$ values including the real axis, with the exception of the specified set of points $\omega_m=(2m+1)\pi T$. In particular, the $\chi(\vec{p},\omega=0)$ value becomes complex. The electron damping on the Fermi surface is determined by the following expression $\operatorname{Im}\Sigma(\omega=0)=\operatorname{Im}\chi(\vec{p},\omega=0)\neq 0$. This term is different from zero, which determines the presence of the damping on the Fermi surface. Moreover, one can say that this damping is increasing with the increase of the temperature as with the temperature increase one moves in the complex plane away from the nearest points $\omega=i\pi T$, and $\omega=-i\pi T$ in which $\chi(\vec{p},\omega)$ is real. With the use of the known property of the Pauli matrixes $\hat{\tau}_3\hat{\tau}_1\hat{\tau}_3=-\hat{\tau}_1$ the following expression for the electron Green's function in the facings of the Pauli matrices is obtained $[27]$:

$$\hat{\tau}_3 g_R(\boldsymbol{p},\omega_m)\hat{\tau}_3=\frac{i\omega_m Z(\vec{p},\omega_m)\hat{\tau}_0+\left[\xi+\chi(\vec{p},\omega_m)\right]\hat{\tau}_3-\varphi(\vec{p},\omega)\hat{\tau}_1.}{\left[i\omega_m Z(\vec{p},\omega_m)\right]^2+\left[\xi+\chi(\vec{p},\omega_m)\right]^2-\varphi^2(\vec{p},\omega).} \tag{2}$$

Using the standard rules of the diagram technique (see, for example, $[27]$) we get for the electron-phonon component of the self-energy part the expression (3)

$$\hat{\Sigma}^{ph}(\boldsymbol{p},\omega_n)=-T\sum_{n'}\int\frac{d^3\boldsymbol{p}'}{(2\pi)^3}\hat{\tau}_3 g(\boldsymbol{p}',\omega_{n'})\hat{\tau}_3\sum_j\left|g_j(\boldsymbol{p},\boldsymbol{p}')\right|^2 D_j(\boldsymbol{p}-\boldsymbol{p}',\omega_n-\omega_{n'}). \tag{3}$$

In (3) $g_j(\boldsymbol{p},\boldsymbol{p}')=-\sum_\kappa\frac{1}{\sqrt{2NM_\kappa\omega_j(\boldsymbol{q})}}\langle\boldsymbol{p}|\boldsymbol{e}_j(\boldsymbol{q},\kappa)\nabla V_{ei\kappa}(\boldsymbol{r})|\boldsymbol{p}'\rangle e^{i\boldsymbol{q}\boldsymbol{\rho}_\kappa}$, where $\omega_j(\boldsymbol{q})$ is a j-branch phonon frequency and $\boldsymbol{e}_j(\boldsymbol{q},\kappa)$ is a polarisation vector corresponding to the impulse $\boldsymbol{q}=\boldsymbol{p}-\boldsymbol{p}'$. Use the following spectral representations for both the electron Green's function

$g(\vec{p},i\omega_m)=\int\limits_{-\infty}^{\infty} dz' a(\boldsymbol{p},z')/(i\omega_m - z')2\pi$ and the phonon Green's function

$D_j(\boldsymbol{p},\omega_n)=\int\limits_{-\infty}^{\infty} dz' b_j(\boldsymbol{p},z')/(i\omega_n - z')2\pi$. Here $a(\boldsymbol{p},z)$ is the spectral density of the thermal electron Green's function, $b_j(\boldsymbol{p}-\boldsymbol{p}',z)$ is the spectral density of a phonon Green's function. The retarded electron Green's function $g_R(\boldsymbol{p},z)$ is related to the spectral density of the thermal electron Green's function $a(\boldsymbol{p},z)$ in the following [37] way: $a(\boldsymbol{p},z)=2\,\mathrm{Im}\,g_R(\boldsymbol{p},z)$. After performance of the analytical continuation $i\omega_p \to \omega + i\delta$ the self-energy part of an electron Green's function of the metal with the use of the following sum:

$$\sum_{n'} \frac{1}{i\omega_n - z'} \frac{1}{i\omega_n - i\omega_{n'} - z} = -\frac{1}{2} \frac{th\frac{z'}{2\mathrm{T}} + cth\frac{z}{2\mathrm{T}}}{i\omega_n - z - z'} \tag{4}$$

shall be given by the following expression:

$$\hat{\Sigma}^{ph}(\boldsymbol{p},\omega)=\int \frac{d^3\boldsymbol{p}'}{(2\pi)^3}\sum_j |g_j(\boldsymbol{p},\boldsymbol{p}')|^2 \int\limits_{-\infty}^{+\infty}\frac{dz}{2\pi}\int\limits_{-\infty}^{+\infty}\frac{dz'}{2\pi}\frac{th\frac{z'}{2\mathrm{T}}+cth\frac{z}{2\mathrm{T}}}{\omega - z - z' + i\delta}\hat{\tau}_3\,\mathrm{Im}\,g_R(\boldsymbol{p}',z')\hat{\tau}_3 b_j(\boldsymbol{p}-\boldsymbol{p}',z). \tag{5}$$

Integration on an impulse shall be represented as follows $\int \frac{d^3\boldsymbol{p}}{(2\pi)^3}\ldots = \int d\xi \int\limits_{S(\xi)} \frac{d^2\boldsymbol{p}}{v_{\boldsymbol{p}}}\ldots$,

where $v_p$ is the electron velocity modulo at the energy surface $\xi$. Introduce a notation for the electron-phonon interaction spectral function $\alpha^2(\xi',\xi,z)F(\xi',\xi,z)$ as follows

$$\alpha^2(\xi',\xi,z)F(\xi',\xi,z)=\frac{1}{2\pi}\int\limits_{S(\xi')}\frac{d^2\boldsymbol{p}'}{v_{\mathbf{p}'}}\sum_j |g_j(\boldsymbol{p},\boldsymbol{p}')|^2 b_j(\boldsymbol{p}-\boldsymbol{p}',z)\left(\int\limits_{S(\xi')}\Gamma(\boldsymbol{p},\boldsymbol{p}')\frac{d^2\boldsymbol{p}'}{v_{\mathbf{p}'}}\right)^{-1}. \tag{6}$$

The phonon contribution to the self-energy part of the electron retarded Green's function $\hat{g}_R$ with the use of the adopted notations shall be expressed in the following form:

$$\hat{\Sigma}^{ph}(\xi,\omega)=\frac{1}{2\pi}\int_{-\infty}^{+\infty}dz\int_{-\infty}^{+\infty}dz'\int_{-\mu}^{+\infty}d\xi'\alpha^2(\xi',\xi,z)F(\xi',\xi,z)N_0(\xi')\frac{th\frac{z'}{2T}+cth\frac{z}{2T}}{\omega-z-z'+i\delta}\hat{\tau}_3\,\mathrm{Im}\,\hat{g}_R(\xi',z')\hat{\tau}_3. \quad (7)$$

Use the technique for the real frequency Éliashberg equations solution. This technique allows us to control the process of calculating $\mathrm{Re}\,Z(\omega)$, $\mathrm{Im}\,Z(\omega)$, $\mathrm{Re}\,\Sigma(\omega)$, $\mathrm{Im}\,\Sigma(\omega)$, $\mathrm{Re}\,\chi(\omega)$ and $\mathrm{Im}\,\chi(\omega)$ frequency behavior and compare the results with the relevant values obtained on the experiment. In (7) $N_0(\xi)$ is a "bare" (not renormalized by the EP interaction) variable electronic density of states defined by the following expression $\int_{S(\xi)}\frac{d^2\boldsymbol{p}'}{v_{\xi\boldsymbol{p}'}}d\xi=\int_{S(\xi)}N_0(\xi)d\xi$ with the energy of the "bare" electrons $\xi$ with the pulse $\boldsymbol{p}$ measured from the Fermi level. It is not assumed that the electron pulses are on the Fermi surface. Neglect in (7) the $\alpha^2F$ dependence on the $\xi$, $\xi'$ variables so that $\alpha^2(\xi',\xi,z)F(\xi',\xi,z)\approx\alpha^2(z)F(z)$. Replace the $Z(\vec{p}',\omega)$ function with $Z(\xi,\omega)$ corresponding to the constant energy $\xi$ in the direction determined by the angle $\varphi$. Average the expression (7) to the angle $\varphi$ of the pulse direction. The phonon contribution to the self energy part $\hat{\Sigma}^{ph}(\xi,\omega)$ (7) of the retarded electron GF $\hat{g}_R$ can be easily reformulated to the following form:

$$\hat{\Sigma}^{ph}(\xi,\omega)=\frac{1}{\pi}\int_{-\infty}^{+\infty}dz'\int_{-\mu}^{+\infty}d\xi'\frac{N_0(\xi')}{N_0(0)}K^{ph}(z',\omega)\hat{\tau}_3\,\mathrm{Im}\,\hat{g}_R(\xi',z')\hat{\tau}_3, \quad (8)$$

where

$$K^{ph}(z',\omega)=\int_0^{+\infty}dz\alpha^2(z)F(z)\frac{1}{2}\left\{\frac{th\frac{z'}{2T}+cth\frac{z}{2T}}{z'+z-\omega-i\delta}-\frac{th\frac{z'}{2T}-cth\frac{z}{2T}}{z'-z-\omega-i\delta}\right\} \quad (9)$$

and the $\delta$ has the infinitesimal positive value. Neglect in (8) the $\hat{\Sigma}^{ph}(\xi,\omega)$ dependence on the $\xi$ variable. Substitute $K^{ph}(z',\omega)$ (9) into (8) and further during the transition from the integration $\int_{-\infty}^{+\infty} dz'$ to the integration $\int_{0}^{+\infty} dz'$ after simple transformations we obtain formulas describing the properties of the normal state of the crystal with a variable density of electronic states. For the normal state the $(1,1)$ component of the matrix SP imaginary part $\operatorname{Im}\Sigma(\omega) = -\operatorname{Im}Z(\omega)\omega + \operatorname{Im}\chi(\omega)$ the following expression with the use of the identities $th\frac{z'}{2\mathrm{T}} = 1 - 2f(z') = -1 + 2f(-z')$, $cth\frac{z}{2\mathrm{T}} = 1 + 2n_B(z)$ may be written down:

$$\operatorname{Im}\Sigma(\omega) = -\pi\int_0^{+\infty} dz\alpha^2(z)F(z)\times$$
$$\times\left\{\left[N(\omega - z) - N(\omega + z)\right]n_B(z) + N(\omega - z)f(z - \omega) + N(\omega + z)f(z + \omega)\right\}. \quad (10)$$

The expression for the real $(1,1)$ component of the self-energy $\hat{\Sigma}^{ph}(\omega)$ matrix part $\operatorname{Re}\Sigma(\omega) = \left[1 - \operatorname{Re}Z(\omega)\right]\omega + \operatorname{Re}\chi(\omega)$ has the following form:

$$\operatorname{Re}\Sigma(\omega) = -P\int_0^{+\infty} dz\alpha^2(z)F(z)\int_0^{+\infty} dz'\left\{\left[n_B(z) + f(-z')\right]\times\right.$$
$$\left.\times\left(-\frac{N(-z')}{z' + z + \omega} + \frac{N(z')}{z' + z - \omega}\right) + \left[n_B(z) + f(z')\right]\left(-\frac{N(-z')}{z' - z + \omega} + \frac{N(z')}{z' - z - \omega}\right)\right\}. \quad (11)$$

In (10) , (11 $n_B(z)$ is the Bose distribution function, $f(z')$ is the Fermi distribution function. In (10), (11) for the bravity $N(z')/N_0(0)$ is marked as $N(z')$. With the help of (10), (11) it is easy to see that $\operatorname{Re}\chi(\omega)$ and $\operatorname{Im}Z(\omega)$ are expressed in terms of the integrals containing the density of states difference at the energies of the opposite sign, and thus reflect a measure of the electron-hole non-equivalence in the EP system, turning to zero at zero electron-hole nonequivalence. In deriving (10), (11) the fact was taken into account that at the temperatures near $T_c$ the anomalous Green's function can be assumed to be equal zero. Given that $cth\left(\frac{\omega_{ph}}{2T}\right) \approx 1$, get as a result the formulae for the normal state, obtained in [38]. The renormalized by the EP interaction the density of the electron states $N(z')$ is expressed through the "bare" density of electron states $N_0(\xi)$

$$N(z') = -\frac{1}{\pi}\int_{-\mu}^{\infty} d\xi' N_0(\xi') \operatorname{Im} g_R(\xi', z'). \qquad (12)$$

Such density of states as $N(z')$ is not the symmetrical (even) function of $z'$. Further during the transition from the integration $\int_{-\infty}^{+\infty} dz'$ to the integration $\int_{0}^{+\infty} dz'$ take into account the $\operatorname{Re} Z(z')$ and $\operatorname{Re}\chi(z')$ parity as well as the oddness property of the functions $\operatorname{Im} Z(z')$ and $\operatorname{Im}\chi(z')$ simultaneously with the following order parameter property $\varphi(-z') = \varphi^*(-z')$ [ 27]. For the superconducting state from (8), taking into account the expression (2) for the retarded electron Green's function $\hat{g}_R(\xi', z')$ we obtain the equations for both the real $\operatorname{Re}\varphi(\omega)$ and the imaginary part $\operatorname{Im}\varphi(\omega)$ of the anomalous Green's function self-energy part as the system of two equations having the following form:

$$\operatorname{Re}\varphi(\omega) = \frac{1}{\pi} P \int_0^{+\infty} dz' \left[K^{ph}(z',\omega) - K^{ph}(-z',\omega)\right] \int_{-\mu}^{+\infty} d\xi' \frac{N_0(\xi')}{N_0(0)} \operatorname{Im} \frac{\varphi(z')}{\left[Z(z')(z')\right]^2 - \varphi^2(z') - \left(\xi' + \chi(z')\right)^2}, \qquad (13)$$

$$\begin{aligned} \operatorname{Im}\varphi(\omega) = \frac{1}{2}\int_0^{+\infty} dz' \{\ & \alpha^2(|\omega - z'|) F(|\omega - z'|)\left[cth\frac{(\omega - z')}{2T} + th\frac{z'}{2T}\right] sign(\omega - z') - \\ & -\alpha^2(|\omega + z'|) F(|\omega + z'|)\left[cth\frac{(\omega + z')}{2T} - th\frac{z'}{2T}\right] sign(\omega + z')\ \} \times \\ & \times \int_{-\mu}^{+\infty} d\xi' \frac{N_0(\xi')}{N_0(0)} \operatorname{Im} \frac{\varphi(z')}{\left[Z(z')(z')\right]^2 - \varphi^2(z') - \left(\xi' + \chi(z')\right)^2} \end{aligned} \qquad (14)$$

Represent complex functions in the following form: $Z(z') = \operatorname{Re} Z(z') + i \operatorname{Im} Z(z')$, $\chi(z') = \operatorname{Re}\chi(z') + i\operatorname{Im}\chi(z')$, $\varphi(z') = \operatorname{Re}\varphi(z') + i\operatorname{Im}\varphi(z')$. By direct calculation of the imaginary part of the complex expression appearing in (13), (14) we get the following formula:

$$\operatorname{Im}\frac{\varphi(z')}{\left[Z(z')(z')\right]^2-\varphi^2(z')-\left(\xi'+\chi(z')\right)^2}=\frac{1}{D}\times$$
$$\times\left\{\operatorname{Im}\varphi(z')\left[\left(\operatorname{Re}Z^2(z')-\operatorname{Im}Z^2(z')\right)z'^2-\left(\xi'+\operatorname{Re}\chi(z')\right)^2+\left(\operatorname{Im}\chi(z')\right)^2-\operatorname{Re}\varphi^2(z')+\operatorname{Im}\varphi^2(z')\right]-\right.$$
$$\left.-2\operatorname{Re}\varphi(z')\left[\operatorname{Re}Z(z')\operatorname{Im}Z(z')z'^2-\operatorname{Im}\chi(z')\left(\xi'+\operatorname{Re}\chi(z')\right)-\operatorname{Re}\varphi(z')\operatorname{Im}\varphi(z')\right]\right\} \tag{15}$$

Here

$$D=\left[\left(\operatorname{Re}Z^2(z')-\operatorname{Im}Z^2(z')\right)z'^2-\left(\xi'+\operatorname{Re}\chi(z')\right)^2+\left(\operatorname{Im}\chi(z')\right)^2-\operatorname{Re}\varphi^2(z')+\operatorname{Im}\varphi^2(z')\right]^2+$$
$$+4\left[\operatorname{Re}Z(z')\operatorname{Im}Z(z')z'^2-\operatorname{Im}\chi(z')\left(\xi'+\operatorname{Re}\chi(z')\right)-\operatorname{Re}\varphi(z')\operatorname{Im}\varphi(z')\right]^2. \tag{16}$$

Neglecting both the $\operatorname{Re}\Sigma(\omega)$ and $\operatorname{Im}\Sigma(\omega)$ dependence on $\xi$ and omitting the small (as will be seen later) quantities $\operatorname{Im}Z(\omega)$, $\operatorname{Re}\chi(\omega)$, $\operatorname{Im}\chi(\omega)$ in (15), (16) we obtain the following non-linear equation for the real part of the complex abnormal order parameter $\varphi$:

$$\operatorname{Re}\varphi(\omega)=-\frac{1}{\pi}P\int_0^{+\infty}dz'\left[K^{ph}(z',\omega)-K^{ph}(-z',\omega)\right]\times$$
$$\times\int_{-\mu}^{+\infty}d\xi'\frac{N_0(\xi')}{N_0(0)}\frac{2\operatorname{Re}\varphi^2(z')\operatorname{Im}\varphi(z')}{\left\{\left[\operatorname{Re}Z(z')\right]^2z'^2-(\xi')^2-\operatorname{Re}\varphi^2(z')+\operatorname{Im}\varphi^2(z')\right\}^2+4\left[\operatorname{Re}\varphi(z')\operatorname{Im}\varphi(z')\right]^2}-$$
$$-\frac{1}{\pi}P\int_0^{+\infty}dz'\left[K^{ph}(z',\omega)-K^{ph}(-z',\omega)\right]\times \tag{17}$$
$$\times\int_{-\mu}^{+\infty}d\xi'\frac{N_0(\xi')}{N_0(0)}\frac{\operatorname{Im}\varphi(z')\left\{\left[\operatorname{Re}Z(z')\right]^2z'^2-(\xi')^2-\operatorname{Re}\varphi^2(z')+\operatorname{Im}\varphi^2(z')\right\}}{\left\{\left[\operatorname{Re}Z(z')\right]^2z'^2-(\xi')^2-\operatorname{Re}\varphi^2(z')+\operatorname{Im}\varphi^2(z')\right\}^2+4\left[\operatorname{Re}\varphi(z')\operatorname{Im}\varphi(z')\right]^2}.$$

For the imaginary part of the complex anomalous order parameter $\varphi$ we similarly obtain the following expression:

$$\operatorname{Im}\varphi(\omega)=-\frac{1}{2}\int_0^{+\infty}dz'\{\ \alpha^2\left(|\omega-z'|\right)F\left(|\omega-z'|\omega-z'\right)\left[cth\frac{(\omega-z')}{2\mathrm{T}}+th\frac{z'}{2\mathrm{T}}\right]sign(\omega-z')-$$
$$-\alpha^2\left(|\omega+z'|\right)F\left(|\omega+z'|\right)\left[cth\frac{(\omega+z')}{2\mathrm{T}}-th\frac{z'}{2\mathrm{T}}\right]sign(\omega+z')\ \}\times$$
$$\times\{\ \int_{-\mu}^{+\infty}d\xi'\frac{N_0(\xi')}{N_0(0)}\frac{2\operatorname{Re}\varphi^2(z')\operatorname{Im}\varphi(z')}{\left\{\left[\operatorname{Re}Z(z')\right]^2z'^2-(\xi')^2-\operatorname{Re}\varphi^2(z')+\operatorname{Im}\varphi^2(z')\right\}^2+4\left[\operatorname{Re}\varphi(z')\operatorname{Im}\varphi(z')\right]^2}+$$
$$+\int_{-\mu}^{+\infty}d\xi'\frac{N_0(\xi')}{N_0(0)}\frac{\operatorname{Im}\varphi(z')\left\{\left[\operatorname{Re}Z(z')\right]^2z'^2-(\xi')^2-\operatorname{Re}\varphi^2(z')+\operatorname{Im}\varphi^2(z')\right\}}{\left\{\left[\operatorname{Re}Z(z')\right]^2z'^2-(\xi')^2-\operatorname{Re}\varphi^2(z')+\operatorname{Im}\varphi^2(z')\right\}^2+4\left[\operatorname{Re}\varphi(z')\operatorname{Im}\varphi(z')\right]^2}\ \}. \tag{18}$$

Near $T_c$ the product $\operatorname{Re}\varphi(z')\operatorname{Im}\varphi(z')$ tends to zero for all values of the argument $z'$, so that the expression

$$\frac{2\operatorname{Re}\varphi(z')\operatorname{Im}\varphi(z')}{\left\{\left[\operatorname{Re}Z(\xi',z')\right]^2z'^2-(\xi')^2-\operatorname{Re}\varphi^2(z')+\operatorname{Im}\varphi^2(z')\right\}^2+4\left[\operatorname{Re}\varphi(z')\operatorname{Im}\varphi(z')\right]^2}$$ becomes delta

function $\pi\delta\left\{\left[\operatorname{Re}Z(z')^2z'^2-(\xi')^2-\operatorname{Re}\varphi^2(z')+\operatorname{Im}\varphi^2(z')\right]^2\right\}\operatorname{sgn}\left[\operatorname{Re}\varphi(z')\operatorname{Im}\varphi(z')\right]$, whereas

the expression $\dfrac{\left\{\left[\operatorname{Re}Z(\xi',z')\right]^2z'^2-(\xi')^2-\operatorname{Re}\varphi^2(z')+\operatorname{Im}\varphi^2(z')\right\}}{\left\{\left[\operatorname{Re}Z(\xi',z')\right]^2z'^2-(\xi')^2-\operatorname{Re}\varphi^2(z')+\operatorname{Im}\varphi^2(z')\right\}^2+4\left[\operatorname{Re}\varphi(z')\operatorname{Im}\varphi(z')\right]^2}$

in the second term of (17), (18) simplifies to the following form:

$\left\{\operatorname{Re}Z(z')^2z'^2-\operatorname{Re}\varphi^2(z')+\operatorname{Im}\varphi^2(z')-\xi'^2\right\}^{-1}$. As a result the nonlinear equation (17) for the real part $\operatorname{Re}\varphi(\omega)$ of the order parameter after $\xi'$ integration with the account of the delta- function properties is taken as follows:

$$\operatorname{Re}\varphi(\omega)=-P\int_0^{+\infty}dz'\left[K^{ph}(z',\omega)-K^{ph}(-z',\omega)\right]\frac{\operatorname{Re}\varphi(z')}{\sqrt{\operatorname{Re}^2Z(z')z'^2-\operatorname{Re}\varphi^2(z')+\operatorname{Im}\varphi^2(z')}}\times$$
$$\times\frac{N_0\left(-\left|\operatorname{Re}^2Z(z')z'^2-\operatorname{Re}\varphi^2(z')+\operatorname{Im}\varphi^2(z')\right|^{\frac{1}{2}}\right)+N_0\left(\left|\operatorname{Re}^2Z(z')z'^2-\operatorname{Re}\varphi^2(z')+\operatorname{Im}\varphi^2(z')\right|^{\frac{1}{2}}\right)}{2N_0(0)}- \tag{19}$$
$$-\frac{1}{\pi}P\int_0^{+\infty}dz'\left[K^{ph}(z',\omega)-K^{ph}(-z',\omega)\right]\int_{-\mu}^{+\infty}d\xi'\frac{N_0(\xi')}{N_0(0)}\frac{\operatorname{Im}\varphi(z')}{\operatorname{Re}Z(z')^2z'^2-\operatorname{Re}\varphi^2(z')+\operatorname{Im}\varphi^2(z')-\xi'^2}.$$

For the imaginary part $\mathrm{Im}\varphi(\omega)$ of the order parameter we obtain as a result of the $\xi$ integration in (18) taking into account the properties of the delta-function the following expression:

$$\mathrm{Im}\varphi(\omega)=-\frac{1}{2}\int_{0}^{+\infty}dz'\Big\{\alpha^2(|\omega-z'|)F(|\omega-z'|)\left[cth\frac{(\omega-z')}{2\mathrm{T}}+th\frac{z'}{2\mathrm{T}}\right]sign(\omega-z')-$$
$$-\alpha^2(|\omega+z'|)F(|\omega+z'|)\left[cth\frac{(\omega+z')}{2\mathrm{T}}-th\frac{z'}{2\mathrm{T}}\right]sign(\omega+z')\Big\}\Bigg\{-\frac{\pi\,\mathrm{Re}\varphi(z')}{\sqrt{\mathrm{Re}^2Z(z')z'^2-\mathrm{Re}\varphi^2(z')+\mathrm{Im}\varphi^2(z')}}\times$$
$$\times\left[\frac{N_0\left(-\left|\mathrm{Re}^2Z(z')z'^2-\mathrm{Re}\varphi^2(z')+\mathrm{Im}\varphi^2(z')\right|^{\frac{1}{2}}\right)}{2N_0(0)}+\frac{N_0\left(\left|\mathrm{Re}^2Z(z')z'^2-\mathrm{Re}\varphi^2(z')+\mathrm{Im}\varphi^2(z')\right|^{\frac{1}{2}}\right)}{2N_0(0)}\right]+$$
$$-P\int_{-\mu}^{+\infty}d\xi'\frac{N_0(\xi')}{N_0(0)}\frac{\mathrm{Im}\varphi(z')}{\left[\mathrm{Re}Z(\xi',z')\right]^2z'^2-(\xi')^2-\mathrm{Re}\varphi^2(z')+\mathrm{Im}\varphi^2(z')}\Bigg\}. \qquad (20)$$

The right hand side of equations (19) - (20) for the complex order parameter $\varphi(\omega)$ contains previously $[17-33]$ not taken into account contributions proportional to $\mathrm{Im}\varphi(z')$. The order parameter will be written in the following form $\varphi(\omega)=\Delta(\omega)|Z(\omega)|$, $|Z(z')|=\left(\mathrm{Re}^2Z(z')+\mathrm{Im}^2Z(z')\right)^{\frac{1}{2}}$, the $z'$ integral in (19), (20) is taken as a principal value, that is marked with the P mark, the negative $-\left|\mathrm{Re}^2Z(z')z'^2-\mathrm{Re}\varphi^2(z')+\mathrm{Im}\varphi^2(z')\right|^{\frac{1}{2}}$ value can not be less than $-\mu$, so that the integration over $z'$ at the negative $z'$ breaks provided $\left|\mathrm{Re}^2Z(z')z'^2-\mathrm{Re}\varphi^2(z')+\mathrm{Im}\varphi^2(z')\right|^{\frac{1}{2}}=\mu$. The integrand is equal to zero under such $z'$ that $\mathrm{Re}^2Z(z')z'^2-\mathrm{Re}\varphi^2(z')+\mathrm{Im}\varphi^2(z')<0$. The root will assume to be positive $\sqrt{\mathrm{Re}^2Z(z')z'^2-\mathrm{Re}\varphi^2(z')+\mathrm{Im}\varphi^2(z')}\geq 0$ for any $z'$ sign. In the wide-band materials such as the metal hydrogen phase in study with the symmetry $I41/AMD$ the logarithmic term in the last two equations with the high accuracy can be set equal to zero, so that the system of equations for the order parameter $\varphi(\omega)$ for the electron-phonon systems with the variable density of electron states takes the form:

$$\operatorname{Re}\varphi(\omega) = -P\int_0^{+\infty} dz'\left[K^{ph}(z',\omega) - K^{ph}(-z',\omega)\right]\frac{\operatorname{Re}\varphi(z')}{\sqrt{\operatorname{Re}^2 Z(z')z'^2 - \operatorname{Re}\varphi^2(z') + \operatorname{Im}\varphi^2(z')}}\times$$

$$\times\frac{N_0\left(-\left|\operatorname{Re}^2 Z(z')z'^2 - \operatorname{Re}\varphi^2(z') + \operatorname{Im}\varphi^2(z')\right|^{\frac{1}{2}}\right) + N_0\left(\left|\operatorname{Re}^2 Z(z')z'^2 - \operatorname{Re}\varphi^2(z') + \operatorname{Im}\varphi^2(z')\right|^{\frac{1}{2}}\right)}{2N_0(0)}. \tag{21}$$

$$\operatorname{Im}\varphi(\omega) = -\frac{1}{2}\int_0^{+\infty} dz'\{\ \alpha^2(|\omega - z'|)F(|\omega - z'|)\left[cth\frac{(\omega - z')}{2\mathrm{T}} + th\frac{z'}{2\mathrm{T}}\right]sign(\omega - z') -$$

$$-\alpha^2(|\omega + z'|)F(|\omega + z'|)\left[cth\frac{(\omega + z')}{2\mathrm{T}} - th\frac{z'}{2\mathrm{T}}\right]sign(\omega + z')\ \}\frac{\pi\operatorname{Re}\varphi(z')}{\sqrt{\operatorname{Re}^2 Z(z')z'^2 - \operatorname{Re}\varphi^2(z') + \operatorname{Im}\varphi^2(z')}}\times$$

$$\times\left[\frac{N_0\left(-\left|\operatorname{Re}^2 Z(z')z'^2 - \operatorname{Re}\varphi^2(z') + \operatorname{Im}\varphi^2(z')\right|^{\frac{1}{2}}\right)}{2N_0(0)} + \frac{N_0\left(\left|\operatorname{Re}^2 Z(z')z'^2 - \operatorname{Re}\varphi^2(z') + \operatorname{Im}\varphi^2(z')\right|^{\frac{1}{2}}\right)}{2N_0(0)}\right]. \tag{22}$$

Assuming the constancy of the "bare" electron density of states $N_0(\omega)$, we can pass from a system of equations (21) - (22) to the conventional system of Éliashberg equations $[17-33]$ in which the final width of the electron band, the pairing outside the Fermi-surface, the variability of the electronic density of states and the effects of electron-hole non-equivalence are all neglected. When assuming constant electronic density of states $N_0\left(\pm\left|\operatorname{Re}^2 Z(z')z'^2 - \operatorname{Re}\varphi^2(z') + \operatorname{Im}\varphi^2(z')\right|^{\frac{1}{2}}\right) = const$ the equations for the complex order parameter $\varphi(\omega)$ near $T_c$ (21)-(22) are simplified to the following form:

$$\operatorname{Re}\varphi(\omega) = -P\int_0^{+\infty} dz'\left[K^{ph}(z',\omega) - K^{ph}(-z',\omega)\right]\frac{\operatorname{Re}\varphi(z')}{\sqrt{\operatorname{Re}^2 Z(z')z'^2 - \operatorname{Re}\varphi^2(z') + \operatorname{Im}\varphi^2(z')}}, \tag{23}$$

$$\operatorname{Im}\varphi(\omega) = -\frac{1}{2}\int_0^{+\infty} dz'\{\ \alpha^2(|\omega - z'|)F(|\omega - z'|)\left[cth\frac{(\omega - z')}{2\mathrm{T}} + th\frac{z'}{2\mathrm{T}}\right]sign(\omega - z') -$$

$$-\alpha^2(|\omega + z'|)F(|\omega + z'|)\left[cth\frac{(\omega + z')}{2\mathrm{T}} - th\frac{z'}{2\mathrm{T}}\right]sign(\omega + z')\ \}\frac{\pi\operatorname{Re}\varphi(z')}{2\sqrt{\operatorname{Re}^2 Z(z')z'^2 - \operatorname{Re}\varphi^2(z') + \operatorname{Im}\varphi^2(z')}}. \tag{24}$$

Usually already extremely oversimplified equations (23) - (24) for the superconducting order parameter are solved neglecting the imaginary part of the order parameter, that leads to the next standard [27] form as follows:

$$\mathrm{Re}\,\varphi(\omega) = -P\int_{0}^{+\infty} dz'\left[K^{ph}(z',\omega) - K^{ph}(-z',\omega)\right]\frac{\mathrm{Re}\,\varphi(z')}{\sqrt{\mathrm{Re}^2 Z(z')z'^2 - \mathrm{Re}\,\varphi^2(z')}}. \quad (25)$$

From the above it is clear that the use of such a simplified form (25) of the Éliashberg equations can not qualify for a quantitative description of the superconducting transition temperature $T_c$ in the specific materials.

## 3. Normal properties of the I41 / AMD metal hydrogen phase at the pressure of 500 GPa

Under homogeneous compression at the pressure of 500 GPa the structure with the symmetry *I*41/*AMD* represents the stable metallic hydrogen phase (Fig.1, [13]). The phonon spectrum of this phase does not contain imaginary frequencies and extends to a maximum frequency of about 340 meV. This structure has a minimal enthalpy among the all calculated phases, but the difference between the enthalpy values of the different phases is small and does not exceed 0.16 eV.

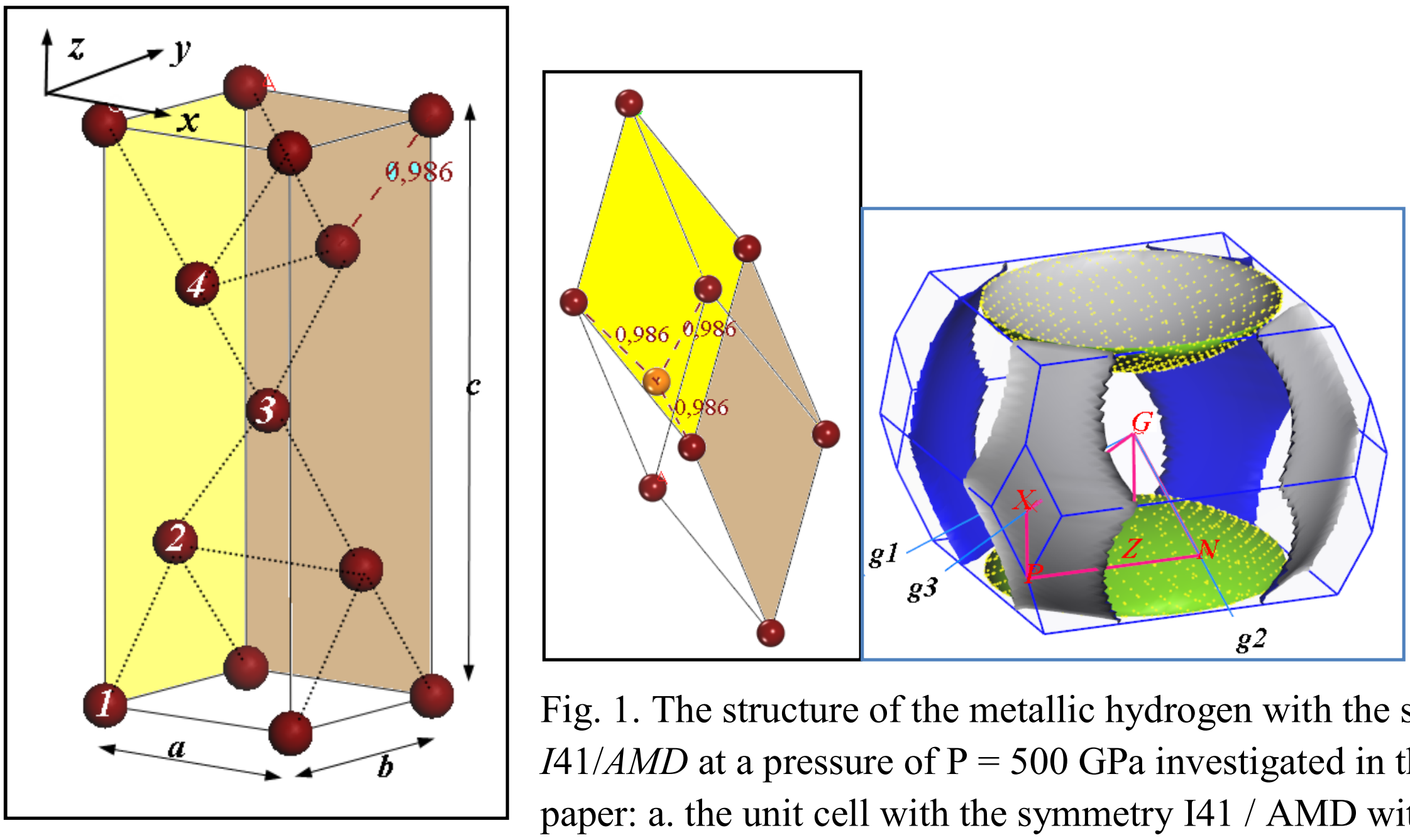

Fig. 1. The structure of the metallic hydrogen with the symmetry *I*41/*AMD* at a pressure of P = 500 GPa investigated in the present paper: a. the unit cell with the symmetry I41 / AMD with a basis of 4 H atoms; b. a primitive cell containing 2 hydrogen H atoms; c. the Fermi surface in the first Brillouin zone of the reciprocal space. The Fig.1 is taken from [13].

To obtain the normal state properties of the *I*41/*AMD* stable metallic hydrogen phase substitute "bare" not renormalized with the electron-phonon interaction electron density of the metal hydrogen states $N_0(\xi')$ (Fig.2.a, [13]) as well as the spectral

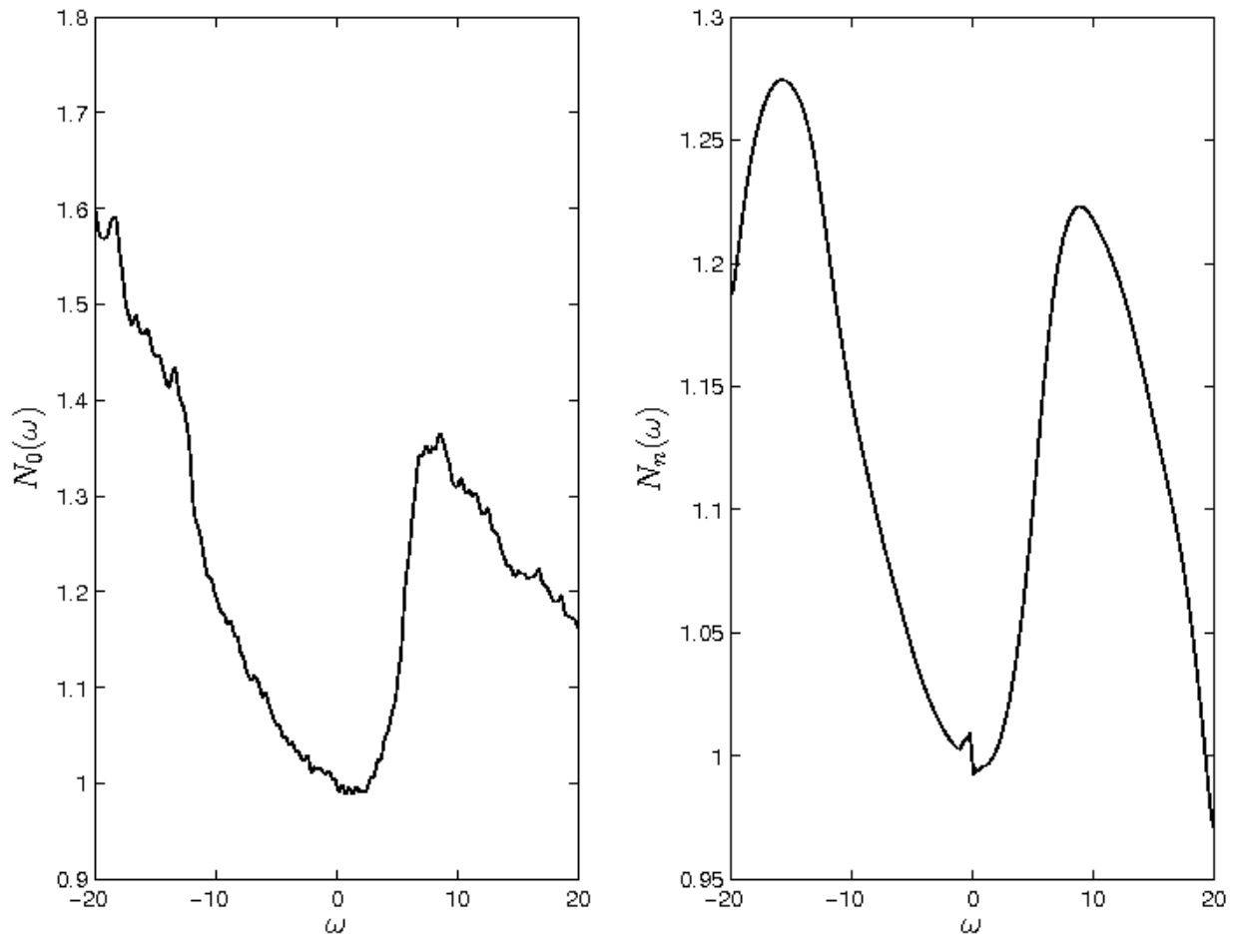

a b

Fig. 2. a. Dimensionless "bare" total density of I41 / AMD metal hydrogen electronic states $N_0$ at the pressure 500 GPa [13]; b. the reconstructed density of I41 / AMD metal hydrogen electronic states $N$ at the pressure 500 GPa. The frequency $\omega$ is expressed in the dimensionless units (as a fraction of the maximum frequency 0.34 eV of the phonon spectrum of the I41 / AMD metal hydrogen phase).

function of the electron-phonon interaction $\alpha^2(z)F(z)$ (Fig. 3) to the nonlinear system of the integral equations (10) - (12), describing the normal metal hydrogen state. $N_0(\xi')$ is a function of a dimensionless argument expressed in the fractions of the maximum phonon frequency. The spectral function of the electron-phonon interaction $\alpha^2(z)F(z)$ (Fig. 3) with the dimensionless argument is also expressed as a function of a fraction of the maximum phonon frequency $0.34eV$ for this metal hydrogen phase in study. To solve the nonlinear system of integral equations (10) - (12) the program of the numerical simulation of the electronic density of states is written. Fig. 4 shows the results of solving the system of equations (10) - (12) for the above mentioned the I41 / AMD metal hydrogen phase for both the real part $\mathrm{Re}\,\Sigma(\omega)$ of the self-energy part of the electron Green's function of the metal hydrogen phase in study and for the imaginary component $\mathrm{Im}\,\Sigma(\omega)$ in the range of the dimensionless energy from -20 to +20. Such a range corresponds to the most interesting energy range in relation to the superconductivity of about $-6.8eV$ to $+6.8eV$ near the Fermi surface. The system of equations (10) - (12) describing the electronic structure of the normal state of the metal hydrogen phase in study with a strong EP interaction is solved at a pressure of $P = 500GPa$ and a temperature $T \sim 200K$ by the method of successive approximations to achieve the self-consistency effect. The behavior of the reconstructed $N(\omega,T)$ (Fig.2. b) electron density of states in this energy range responds to the presence of the major contributions to the density of states in the reconstructed conduction band. Such a behavior of the electron density of states is crucial for the existence of the superconducting properties of the material. The Fermi level crosses two bands and is close to the minimum density of electronic states. This fact points to the need to find more favorable for the superconductivity of the metallic hydrogen stable phases, in which the Fermi level would have been at the peak of the density of electronic states.

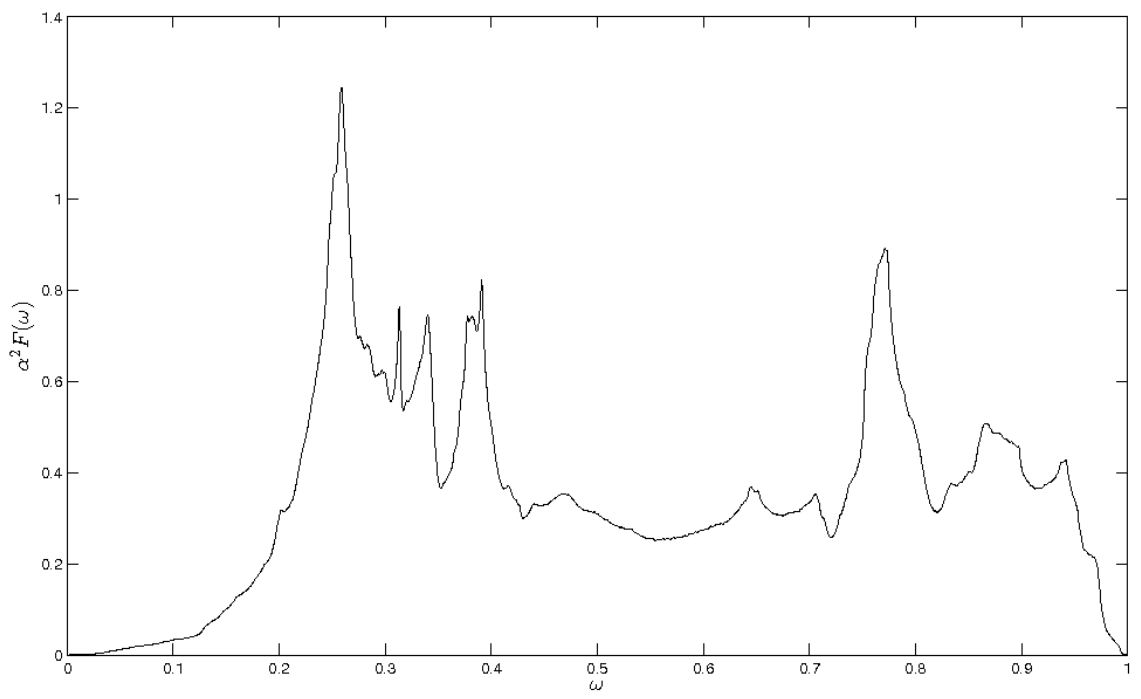

Fig. 3. The spectral function of the electron-phonon interaction $\alpha^2(\omega)F(\omega)$ in the I41 / AMD metal hydrogen phase at the pressure 500 GPa. The frequency $\omega$ is expressed in the dimensionless units (as a fraction of the maximum frequency 0.34 eV of the phonon spectrum of the I41 / AMD metal hydrogen phase).

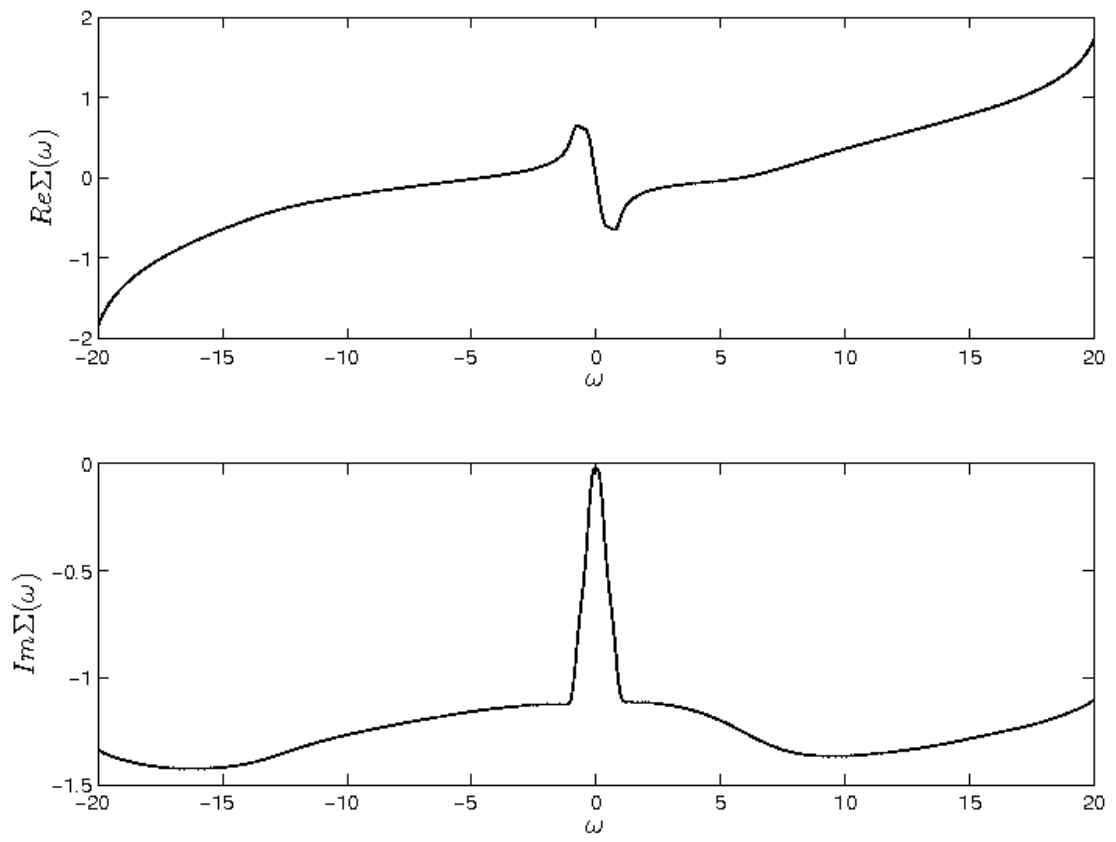

Fig. 4. The real part $\mathrm{Re}\,\Sigma(\omega)$ (up) and the imaginary part $\mathrm{Im}\,\Sigma(\omega)$ of the self-energy part of the electron Green's function (down) in the I41 / AMD metal hydrogen phase at a pressure of $P = 500\,GPa$ and at the temperature $T = 200\,K$. Both graphs are shown in the range of 6.8 electron-volts at the both sides of the Fermi level. The frequency is expressed in the dimensionless units as a fraction of the maximum frequency of the phonon spectrum for the I41 / AMD metal hydrogen phase in study being equal to 0.34 eV.

The functional $Z(\omega,T)$, $\chi(\omega,T)$, $N(\omega,T)$ dependencies at the different temperatures, contained in the expressions (10) - (11), were calculated using the developed formalism (see also$[38,39]$). The frequency dependence of both the electron mass operator $\mathrm{Re}\,Z(\omega,T)$, $\mathrm{Im}\,Z(\omega,T)$ and of the terms $\mathrm{Re}\,\chi(\omega)$, $\mathrm{Im}\,\chi(\omega)$ usually called the " complex renormalization of the chemical potential" , are all presented in Fig.5:

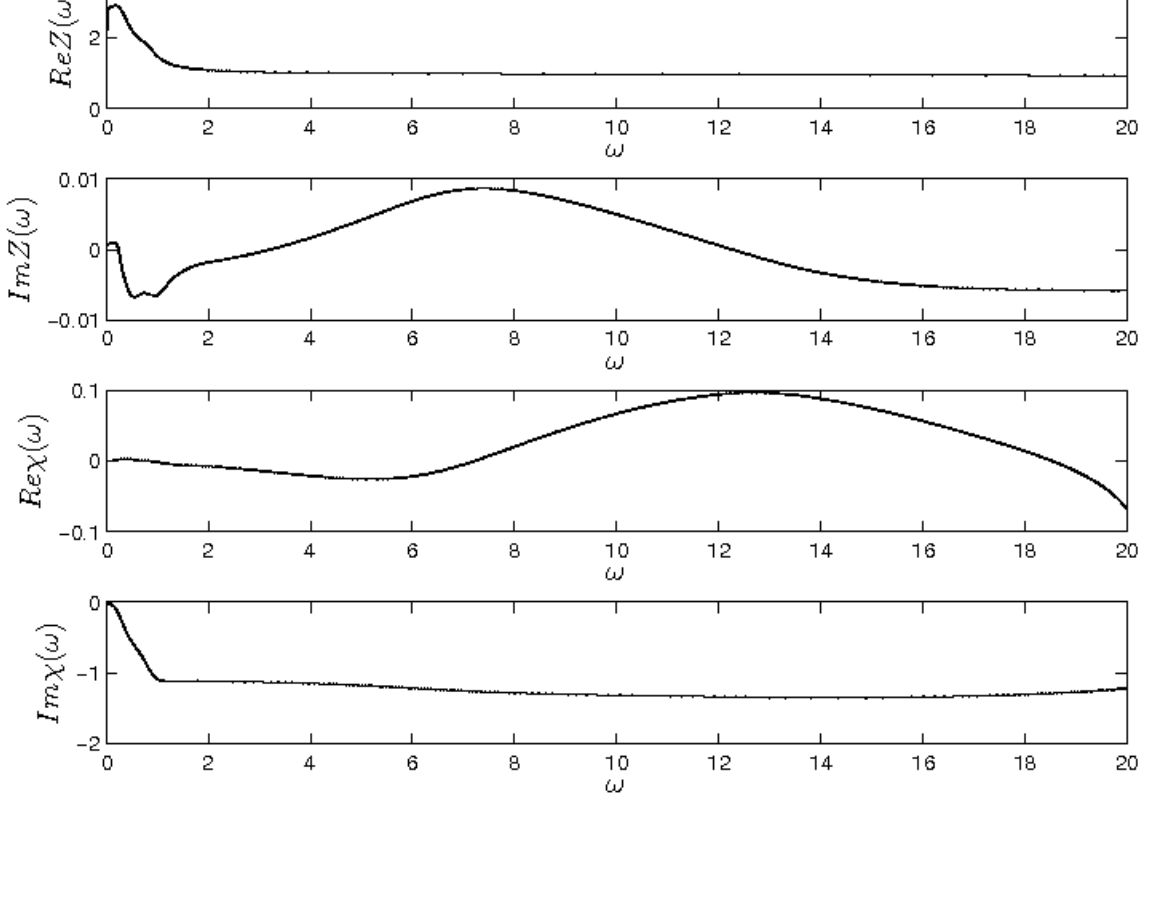

Fig. 5. The reconstructed conductivity band parameters of the I41 / AMD metal hydrogen phase. From up to down: a. The real part $\operatorname{Re} Z(\omega)$ of the electron mass renormalization $Z(\omega)$, b. The imaginary part $\operatorname{Im} Z(\omega)$ of the renormalization of the electron mass in the self-energy part of the electron Green's function. c. Renormalized by the electron-phonon interaction the real part $\operatorname{Re}\chi(\omega)$ of the renormalization of the chemical potential. d. Renormalized by the electron-phonon interaction the imaginary part of the renormalization of the chemical potential $\operatorname{Im}\chi(\omega)$. Frequency $\omega$ is expressed in the dimensionless units (as a fraction of the maximum frequency 0.34 eV of the phonon spectrum for the metal hydrogen phase in study). All results were obtained for the pressure P = 500 GPa and the temperature T = 215K.

## 4.High $T_c$ in the metal hydrogen as a result of the variability of the density of the electron states in the band

In the present paper we determine $T_c$ and the frequency behavior of the complex order parameter $\varphi(\omega)$ at the different temperatures by solving the Éliashberg equations in the form of the non-linear system of equations (21) - (22) for the complex order parameter $\varphi(\omega)$ in the metal hydrogen I41 / AMD phase (Fig. 1) at the pressure of 500 GPa with the full account of the variable character of the electronic density of states (Fig.2). In (21)-(22) we have neglected the Coulomb contribution to the order parameter in view of the smallness of the Coulomb pseudopotential $\mu^* \approx 0.1$ in the metal hydrogen I41/AMD phase compared with the significant calculated constant $\lambda = 2\int_0^\infty d\omega \frac{\alpha^2(\omega)F(\omega)}{\omega} \approx 1.68$ of the EP interaction in the metal hydrogen phase. The behavior of the $K^{ph}(z',\omega)$-function (9) for the I41 / AMD metal hydrogen phase at the pressure of 500 GPa calculated with the spectral function of the electron-phonon interaction $\alpha^2(\omega)F(\omega)$ (Fig.2) is shown in Fig. 6:

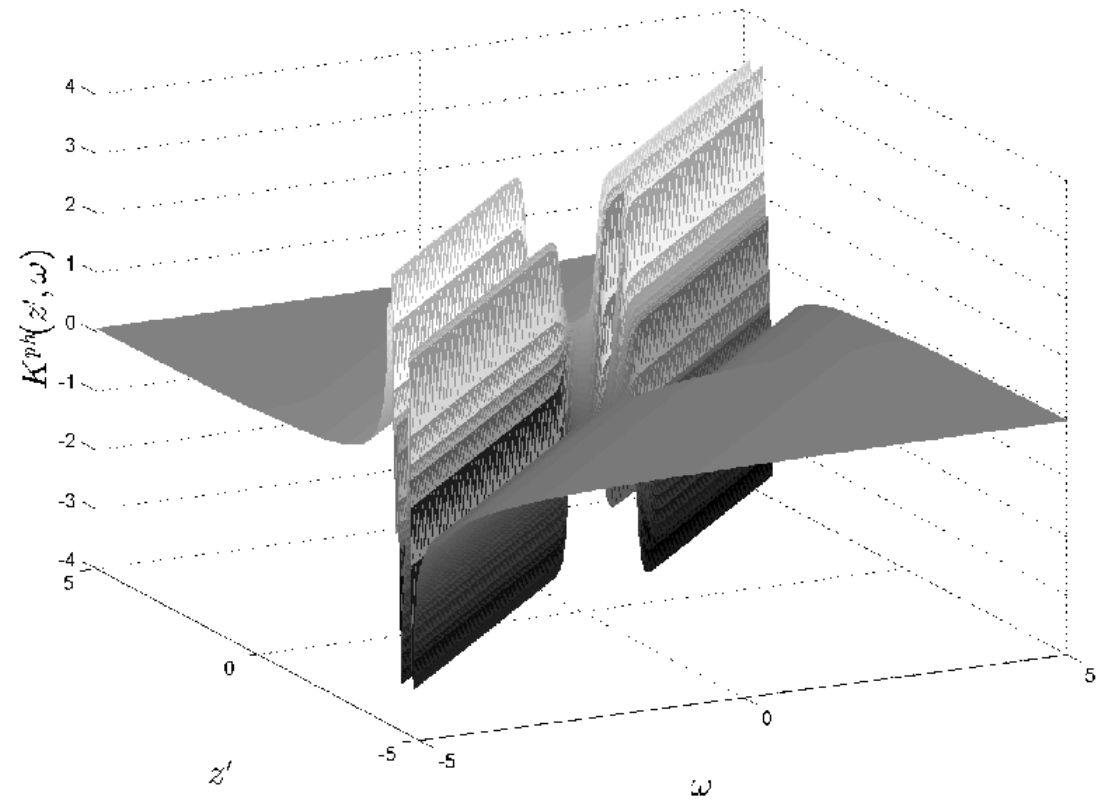

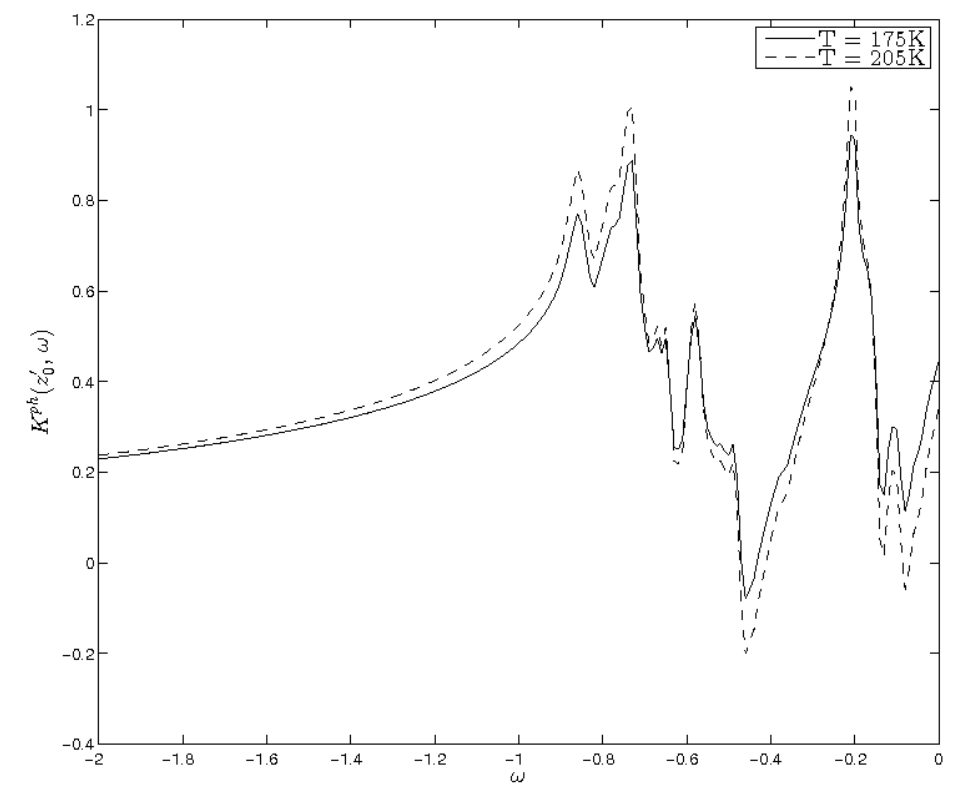

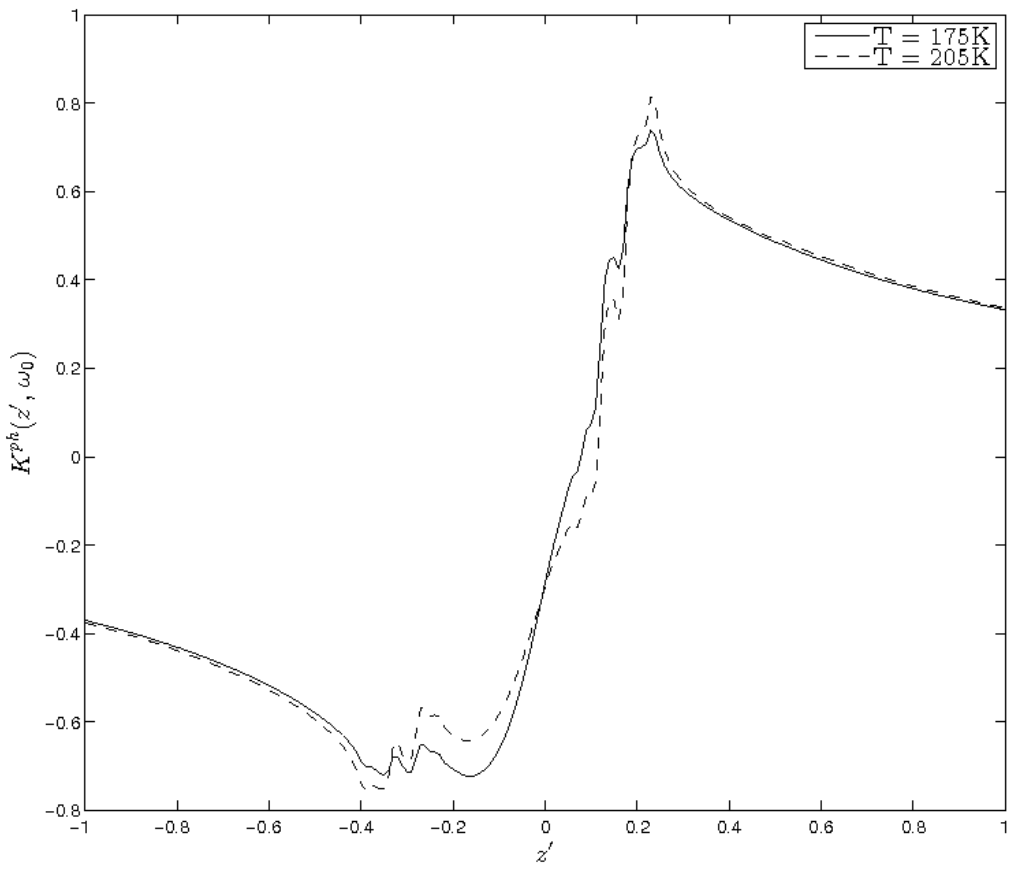

Fig. 6 a. K-function (9) of the I41 / AMD metal hydrogen phase as a function of two dimensionless parameters $z'$ and $\omega$ at the temperature T=205K at the pressure of 500 GPa; The comparative behavior of the $K^{ph}(z',\omega)$ graphs at T=175K and at T = 205K for two cases: b. fixed $z' = 0.11$ value, $\omega$ varies on the interval [-2, 0] c. fixed value $\omega = -0.08$, $z'$ is changed in the interval [-1;1].

The system of equations (21), (22) is solved by using an iterative method based on the frequency behavior of the spectral function of the EP interaction (Éliashberg function) $\alpha^2 F(z)$ (Fig. 3) for the I41 / AMD metal hydrogen phase at the pressure P = 500 GPa. It is found that the process of the convergence of the solution for the real part of the order parameter $\mathrm{Re}\,\varphi(\omega)$ in solving the system of equations (21), (22) is set when the number of 96 of iterations is reached. At T = 217 K $\mathrm{Re}\,\varphi(\omega)$ as well as $\mathrm{Im}\,\varphi(\omega)$ tends to zero with increasing the number of iterations thus indicating no effect of superconductivity at this temperature. In this case, however, the order parameter preserves the characteristic structure of the superconducting state decreasing with the increasing number of iterations. Equations (21), (22) below the $T_c$ temperature have in the case of the presence of the superconductivity a set of three solutions namely: $\mathrm{Re}\,\varphi(\omega)$ and $\mathrm{Im}\,\varphi(\omega)$, $-\mathrm{Re}\,\varphi(\omega)$ and $-\mathrm{Im}\,\varphi(\omega)$ and additionally the zero unstable solution. In the numerical solution of equations (21) - (22) on the real axis the solution before the establishment of the final solution undergoes multiple rebuilds from the "negative" to the "positive" solutions. An additional difficulty in solving the equations (21) - (22) is a numerical integration of improper integrals with divergences appearing in these equations. The behavior of the real part of the order parameter $\mathrm{Re}\,\varphi(\omega)$ as well as the behavior of the imaginary part of the order parameter $\mathrm{Im}\,\varphi(\omega)$ at the temperatures T = 185K, 195K, 205K, 215K is shown in Fig. 7.

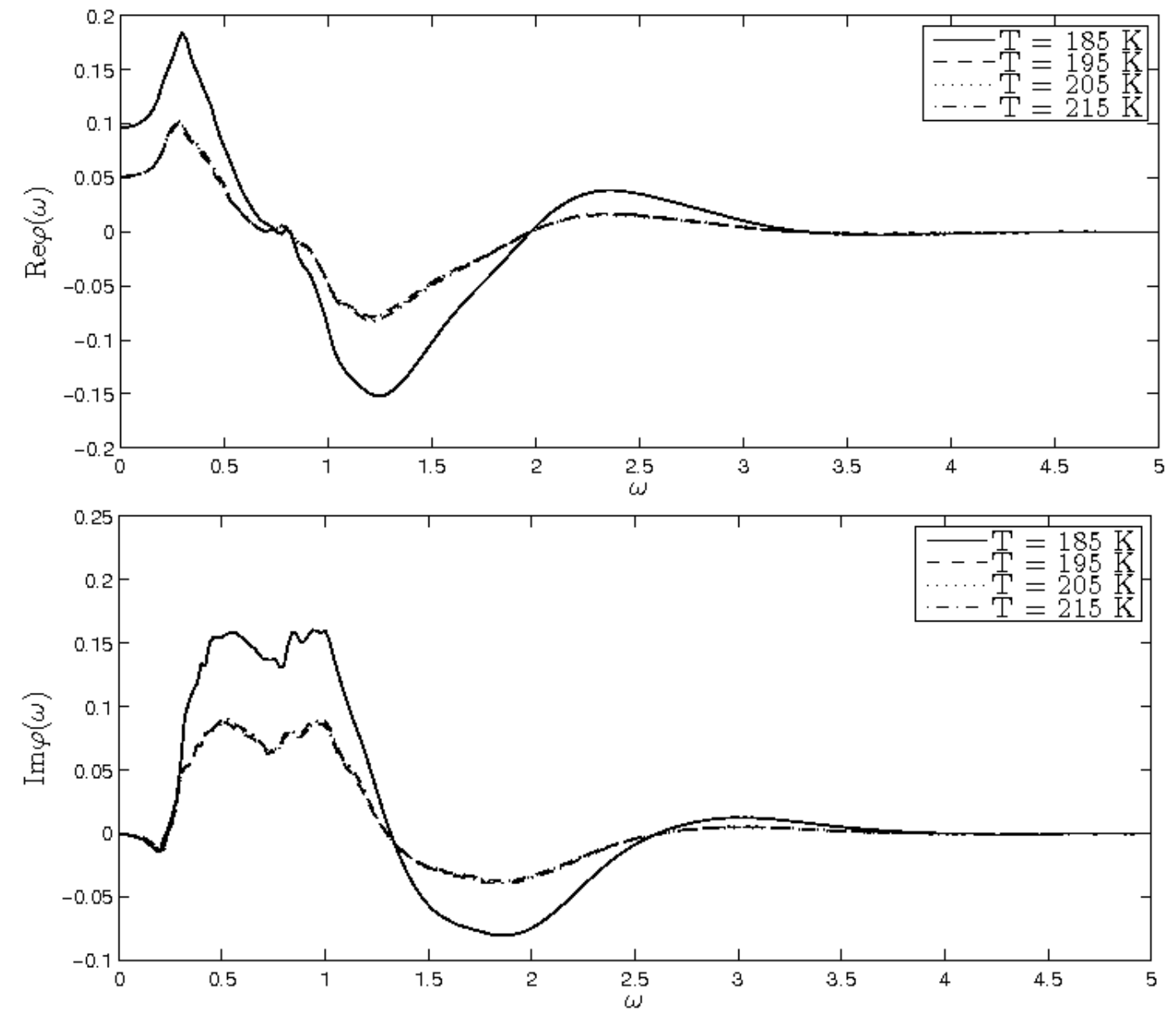

Fig. 7. The frequency dependence of the steady solution for I41 / AMD metal hydrogen phase for a. the real part $\mathrm{Re}\,\varphi(\omega)$ and b. the imaginary part $\mathrm{Im}\,\varphi(\omega)$ of the metal hydrogen phase order parameter at T =185K, 195K, 205K, 215K and at the pressure P = 500 GPa. The frequency is expressed in the dimensionless units, corresponding to the limiting frequency $340\ meV$ of the phonon spectrum for the I41 / AMD metal hydrogen phase.

The detailed behavour of order parameter solution for the temperature T = 215K near the $T_c$ is presented at the Fig.8.

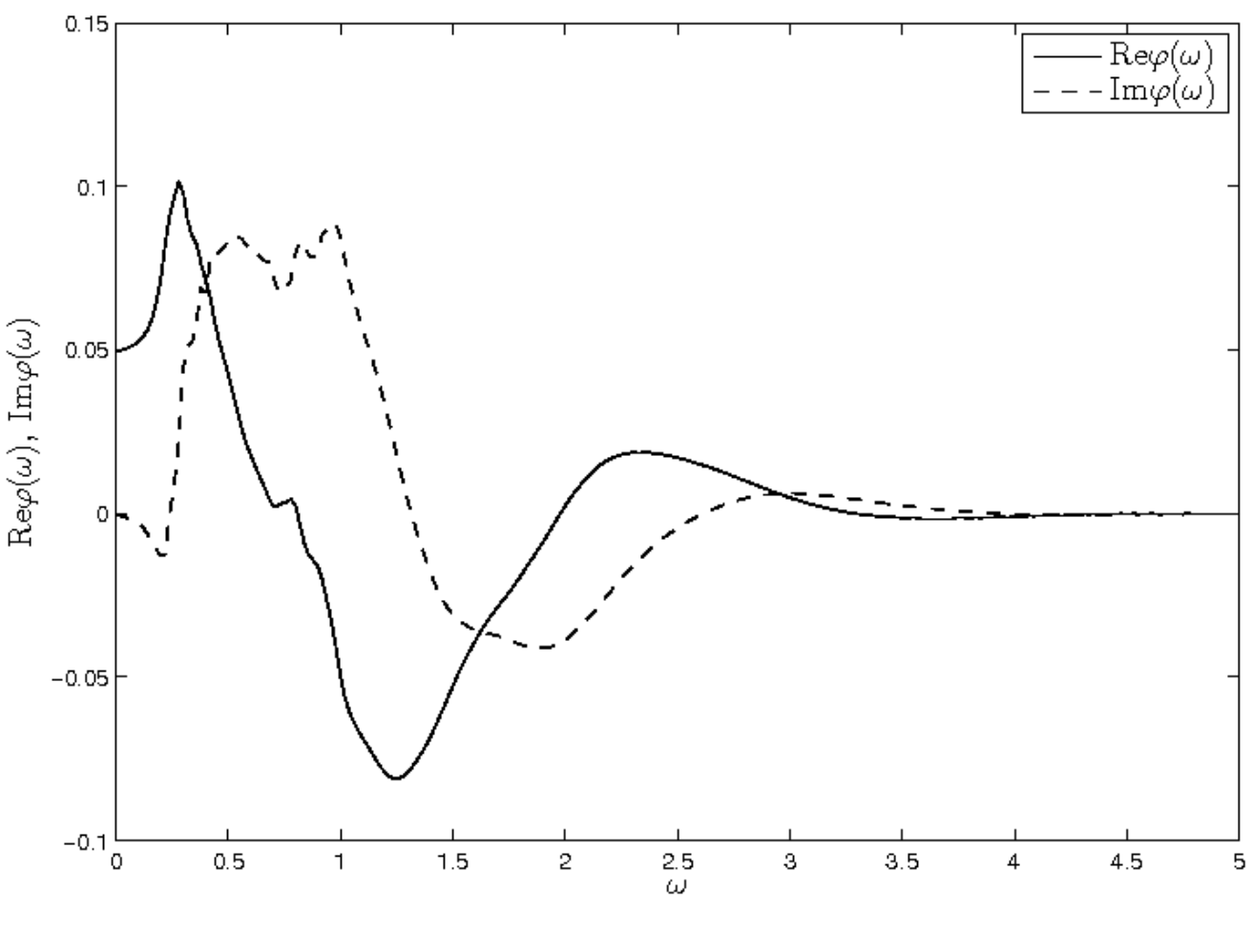

Fig. 8. The frequency dependence of the steady solution for the I41 / AMD metal hydrogen phase for both the real part $\mathrm{Re}\,\varphi(\omega)$ and the imaginary part $\mathrm{Im}\,\varphi(\omega)$ of the I41 / AMD metal hydrogen phase order parameter at the temperature T = 215K and at the pressure P = 500 GPa. The frequency is expressed in the dimensionless units, corresponding to the limiting frequency $340\ meV$ of the phonon spectrum.

The order parameter solution for the resulting value of $T_c$ = 217K is not presented here due to the vanishing order parameter values at this temperature. The imaginary part $\mathrm{Im}\,\varphi(\omega)$ of the order parameter at low frequency is negative, while at the value of the dimensionless frequency equal to 0.23 (Fig.8.) the imaginary part $\mathrm{Im}\,\varphi(\omega)$ value becomes positive. Thus, we set the value of the energy gap in the I41 / AMD metal hydrogen phase, which turned out to be $0.23\times 360 meV$, that is, $83 meV$.

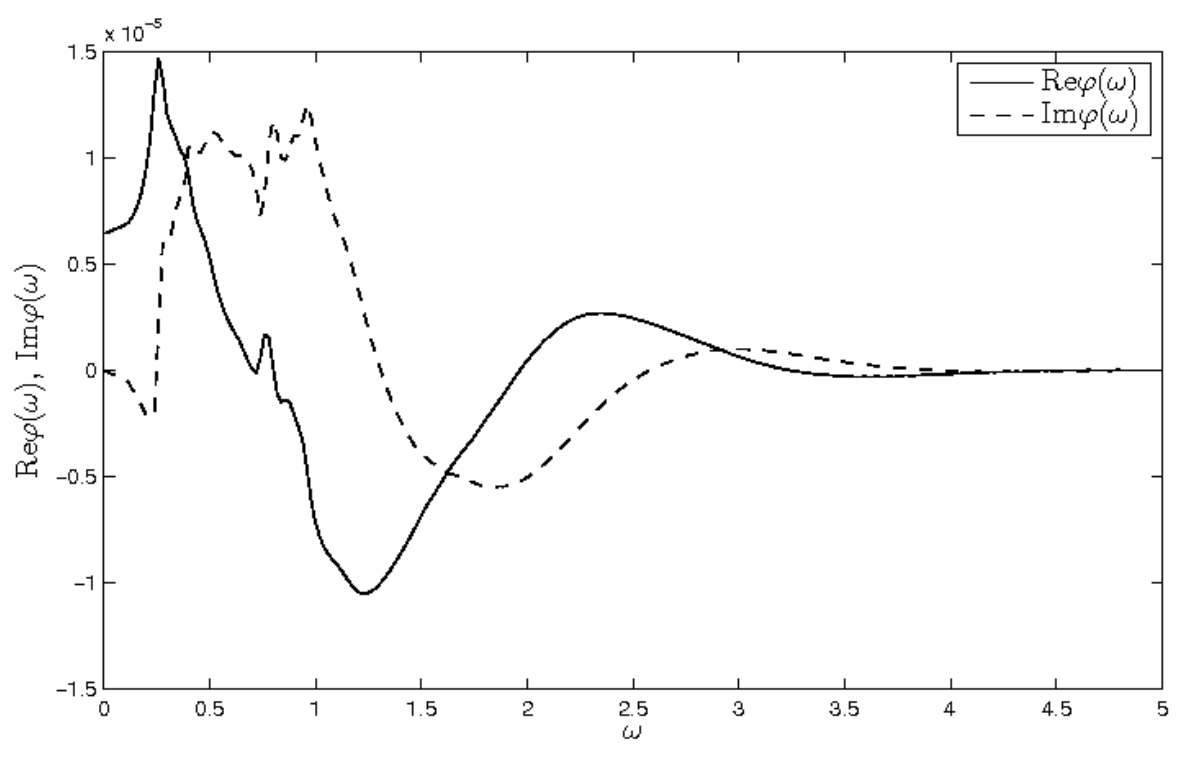

Fig. 9. The frequency dependence of the order parameter solution for the I41 / AMD metal hydrogen phase at the 96th iteration for both the real part $\mathrm{Re}\,\varphi(\omega)$ and the imaginary part $\mathrm{Im}\,\varphi(\omega)$ at T = 220 K and at the pressure P = 500 GPa. The frequency is expressed in the dimensionless units, corresponding to the limiting frequency $340 meV$ of the phonon spectrum.

Fig. 9 shows the frequency dependence at the number 96 of iterations of the very small quantities $\mathrm{Re}\,\varphi(\omega)$, $\mathrm{Im}\,\varphi(\omega)$ at the temperature T = 220K exceeding $T_c$. So, Fig. 9 shows the the extremely slow character of the order parameter $\varphi(\omega)$ vanishing with the increasing number of iterations for the solution at the temperature T = 220K, thus indicating that $T_c < 220K$. From the Fig. 9 it is clearly seen that the order parameter $\mathrm{Re}\,\varphi(\omega)$, $\mathrm{Im}\,\varphi(\omega)$ magnitude in the decrease with the increasing number of iterations even at T = 220K retains characteristic functional dependency of the metal hydrogen order parameter in the superconducting state. At the same time the rough approximations of 1 for $\mathrm{Re}\,\varphi(\omega)$ and of 0 for $\mathrm{Im}\,\varphi(\omega)$ are used as the initial conditions introducing in the solution no functional dependency on the frequency $\omega$.

## 5. Conclusions

Analyzing the results and summing everything written before, we arrive at the following conclusions: 1. Éliashberg theory is generalized to account for the variable nature of the electron density of states. 2. The mathematical method for the solving of Éliashberg equations on the real axis is developed. 3. The generalized Éliashberg equations with the variable nature of the electron density of states are solved. The $T_c = 217K$ prediction for the future experiment for the I41 / AMD metal hydrogen phase is obtained. The fact of the very slow convergence of the solutions of Éliashberg equations with increasing the iteration number is established. 4. The frequency dependence as well as the fine structure of both the real part $\mathrm{Re}\,\varphi(\omega)$ and the imaginary part $\mathrm{Im}\,\varphi(\omega)$ of the order parameter corresponding to the selected I41 / AMD metal hydrogen phase at the temperatures T=185K, 195K, 205K, 215K is obtained. The variation of the frequency dependence of the order parameter with the temperature is presented. 5. The magnitude of the energy gap of the I41 / AMD metal hydrogen phase is found to have the $83meV$ value. 6. We got $T_c = 217K$ for the I41 / AMD metal hydrogen phase. It is shown that at the temperatures above the critical temperature $T_c = 217K$ the order parameter very slowly tends to zero with increasing the number of iterations, maintaining the characteristic functional behavior of the superconducting state on the frequency. At the temperature $T = 220K > T_c$, when the equation for the order parameter leads to an extremely small maximum values of the order parameter $\mathrm{Re}\,\varphi,\ \mathrm{Im}\,\varphi \sim 10^{-8}$ on the hundredth iteration, the order parameter frequency dependence is similar with the order parameter dependence on the frequency for the superconducting state. 7. The method of the solving of the Éliashberg equations on a set of discrete points on the imaginary axis, faced with the problem of the convergence of the solution at the low order of the discrete matrix, does not accurately reproduce the dependence of the order parameter on the frequency after the analytic continuation of the order parameter to the real frequency axis compared to the method of solving the Éliashberg equations on the real axis. 8. All the calculations were carried out ab-initio. The work is not using any assumptions or any fitting parameters. The entire treatment was carried out on the real axis so as to be able to explore the frequency behavior of the order parameter simultaneously with the $T_c$ calculation. 9. Three factors influence the $T_c$ of the EP system of the I41 / AMD metal hydrogen phase in the critical mode, namely: the variable nature of the electron density of states $N_0(\varepsilon)$, the specific properties of a substance depending on the electron mass renormalization $\mathrm{Re}\,Z(z)$, the damping $\mathrm{Im}\,Z(z)$ of the electrons, both the real $\mathrm{Re}\,\chi(\omega)$ and the imaginary $\mathrm{Im}\,\chi(\omega)$ components of the renormalization of the chemical potential, as well as the novel contributions $\sim \mathrm{Im}\,\varphi(\omega)$ to the Éliashberg equations which were not taken into account in the formalism of the previous works $[17-33]$.The neglect of the terms proportional to $\mathrm{Im}\,\varphi(\omega)$ leads to the violation of the Kramers-Kronig relations for the imaginary and the real part of the order parameter in the Éliashberg equations. 10. It is of critical importance to take into account the variability of the density of electron states in the conduction band for the high $T_c$ appearance in the EP system for the I41 / AMD metal hydrogen phase. The accounting of the electron density of states variability in the band leads to the possibility of the pairing of electrons in the entire Fermi – volume in contrast to the usually considered pairing within the layer with $\omega_D$ thickness at the Fermi surface.11. The Coulomb pseudopotential of electrons in the metal hydrogen leads to the inessential $T_c$ reduction. The $T_c$ value in the EP system in the metal hydrogen phase can be greatly improved as compared to the theoretically determined $T_c = 217K$ value in the metal hydrogen by changing the pressure with the simultaneous selection of the optimal $\mathrm{Re}\,Z(\omega)$, $\mathrm{Im}\,Z(\omega)$, $\mathrm{Re}\,\chi(\omega)$, $\mathrm{Im}\,\chi(\omega)$ behavior along with the optimal behavior of the electron density of states $N_0(\varepsilon)$ with a moderate value of the EP interaction strength.

The study was supported by a grant from the Russian Science Foundation (project №14-11-00258).

[a] Corresponding author: *EAMazur@mephi.ru*